\documentclass[twocolumn,tighten]{aastex61}

\usepackage{graphicx}
\usepackage{ulem}
\usepackage{amsmath}
\usepackage{wrapfig}

\begin{document}
\shorttitle{Modeling NGC 3201}
\shortauthors{Kremer et al.}

\title{How black holes shape Globular Clusters: Modeling NGC 3201}
\author[0000-0002-4086-3180]{Kyle Kremer}
\affil{ Department of Physics \& Astronomy, Northwestern University, Evanston, IL 60202, USA}
\affil{ Center for Interdisciplinary Exploration \& Research in Astrophysics (CIERA), Evanston, IL 60202, USA}
\author {Claire S. Ye}
\affil{ Department of Physics \& Astronomy, Northwestern University, Evanston, IL 60202, USA}
\affil{ Center for Interdisciplinary Exploration \& Research in Astrophysics (CIERA), Evanston, IL 60202, USA}
\author[0000-0002-3680-2684]{Sourav Chatterjee}
\affil{ Department of Physics \& Astronomy, Northwestern University, Evanston, IL 60202, USA}
\affil{ Center for Interdisciplinary Exploration \& Research in Astrophysics (CIERA), Evanston, IL 60202, USA}
\affil{Tata Institute of Fundamental Research, Homi Bhabha Road, Mumbai 400005, India}
\author{Carl L. Rodriguez}
\affil{MIT-Kavli Institute for Astrophysics and Space Research, Cambridge, MA 02139, USA}
\author[0000-0002-7132-418X]{Frederic A. Rasio}
\affil{ Department of Physics \& Astronomy, Northwestern University, Evanston, IL 60202, USA}
\affil{ Center for Interdisciplinary Exploration \& Research in Astrophysics (CIERA), Evanston, IL 60202, USA}

\begin{abstract}
Numerical simulations have shown that black holes (BHs) can strongly influence the evolution and present-day observational properties of globular clusters (GCs). Using a Monte Carlo code, we construct GC models that match the Milky Way (MW) cluster NGC 3201, the first cluster in which a stellar-mass BH was identified through radial-velocity measurements. We predict that NGC 3201 contains $\gtrsim 200$ stellar-mass BHs. Furthermore, we explore the dynamical formation of main sequence--BH binaries and demonstrate that systems similar to the observed BH binary in NGC 3201 are produced naturally. Additionally, our models predict the existence of bright blue-straggler--BH binaries unique to core-collapsed clusters, which otherwise retain few BHs.

\end{abstract}

\section{Introduction}
\label{sec:intro}

In recent years, an increasing number of stellar-mass black hole (BH) candidates have been observed within globular clusters (GCs) as members of binary systems with luminous companions. The first such systems were observed in mass-transferring configurations and identified through radio and/or X-ray observations. To date, mass-transferring binaries with BH-candidate accretors have been identified in four Milky Way (MW) GCs: M10 \citep{Strader2014}, M22 \citep[two sources;][]{Strader2012}, 47 Tuc \citep{Miller-Jones2014}, and M62 \citep{Chomiuk2013}.  Additionally, multiple candidates have been identified in extragalactic GCs \citep[e.g.,][]{Maccarone2007,Irwin2010}.

Most recently, a stellar-mass BH candidate was identified in the MW GC NGC 3201 through radial-velocity observations in a detached binary with a main-sequence (MS) companion \citep{Giesers2018}. This observation marks the first detection of a stellar-mass BH candidate through radial-velocity methods.

The effect of BHs on the evolution of GCs is of high current interest, especially since the recent discoveries of merging binary BHs (BBHs) by LIGO \citep{Abbott2016a,Abbott2016b,Abbott2016c,Abbott2016d,Abbott2016e, Abbott2017}. Several recent studies have demonstrated that dynamical processing in GCs can be a dominant formation channel for the BBHs observed by LIGO \citep[e.g.,][]{Banerjee2010,Ziosi2014,Rodriguez2015,Rodriguez2016,Chatterjee2017}. Furthermore, numerical simulations have shown that the retention fraction of BHs in GCs can significantly influence the way the host cluster evolves \citep[e.g.,][]{Mackey2008,Morscher2015,Chatterjee2017}, which in turn affects the overall BH dynamics and BBH formation. 

The observation of stellar-mass BH candidates in GCs has sparked extensive theoretical study of the dynamical formation of BH--non-BH binaries in GCs. Traditionally, it has been argued that finding even a few BH--binary candidates in GCs indicates a much larger population of undetectable BHs retained in those clusters \citep[e.g.,][]{Strader2012,Umbreit2012}. However, recent analyses \citep[e.g.,][]{Chatterjee2017,Kremer2018} have shown that while BH--non-BH binaries can readily form in GCs in both detached and mass-transferring configurations, the number of BH--non-BH binaries retained within a cluster at late times is independent of the total number of BHs retained.

In this letter, we use our Cluster Monte Carlo code (\texttt{CMC}) to construct a GC model that matches the observed properties of NGC 3201, with the goal of exploring the formation of detached BH--non-BH binaries in such a cluster. We demonstrate a large population of BHs ($\gtrsim 200$) is necessary to produce a GC similar to NGC 3201. Furthermore, we show that our GC models that are most similar to NGC 3201 contain BH--MS binaries with orbital parameters similar to those of the system recently observed in NGC 3201.

In Section \ref{sec:method}, we describe our method for modeling GCs and introduce the grid of models used for this analysis. In Section \ref{sec:results}, we display various observational properties of our models and identify the models which best resemble the observed properties of NGC 3201. In Section \ref{sec:MS}, we discuss the properties and dynamical history of BH--MS binaries found in our models at late times and compare these systems to the BH--MS recently identified in NGC 3201. In Section \ref{sec:BS}, we discuss the formation of blue straggler--BH binaries found in our core-collapsed GC models which retain few BHs. We discuss our results and conclude in Section \ref{sec:discussion}.

\section{Method}
\label{sec:method}

\subsection{Globular cluster models}

In order to model the long-term evolution of GCs, we use \texttt{CMC} \citep{Joshi2000,Joshi2001, Fregeau2003, Umbreit2012, Pattabiraman2013, Chatterjee2010, Chatterjee2013a,Rodriguez2018}. \texttt{CMC} incorporates all physics relevant to the evolution of BHs and BH binaries in GCs, including two-body relaxation \citep{Henon1971a, Henon1971b}, single and binary star evolution \citep[calculated using updated versions of the \texttt{SSE} and \texttt{BSE} packages;][]{Hurley2000,Hurley2002,Chatterjee2010}, three-body binary formation \citep{Morscher2013}, galactic tides \citep{Chatterjee2010}, and small-$N$ gravitational encounters calculated using \texttt{Fewbody} \citep{Fregeau2004,Fregeau2007}, updated to incorporate post-Newtonian effects in all three- and four-body encounters \citep{Rodriguez2018, AmaroSeoane2016, Antognini2014}.

 
 Here, we present a grid of 16 GC models aimed at representing a cluster like NGC 3201. Fixed initial cluster parameters include total particle number ($N=8 \times 10^5$), virial radius ($r_v = 1$ pc), King concentration parameter ($W_0=5$), binary fraction ($f_b = 5$\%), and metallicity ($Z=0.001$).

The initial stellar masses are sampled from the initial mass function (IMF) given in \citet{Kroupa2001} in the range 0.08--150 $M_{\odot}$. 

The initial number of binary stars is chosen based on the specified $f_b$ and $N$. Primary masses for all binaries are sampled from the same IMF as single stars, and secondary star masses are assigned assuming a flat distribution in mass ratios. As in previous papers \citep[e.g.,][]{Kremer2018}, initial orbital periods, $P$, for all binaries are drawn from a distribution of the form $dn/d\log P\propto P^0$ and the initial eccentricities are thermal.
 
To treat stellar remnant formation, we use a modified prescription from that implemented in \texttt{SSE} and \texttt{BSE} by using the results of \citet{Fryer2001} and \citet{Belczynski2002}. Upon formation, all neutron star (NS) remnants receive natal kicks drawn from a Maxwellian distribution with $\sigma_{\rm{NS}} = 265\,\rm{km\,s}^{-1}$ \citep{Hobbs2005}. For BH natal kicks, we assume kick magnitudes independent of the BH masses and drawn from a Maxwellian with dispersion width, $\sigma_{\rm{BH}}$. The ratio $\sigma_{\rm{BH}}/\sigma_{\rm{NS}}$ is varied between models from $\sigma_{\rm{BH}}/\sigma_{\rm{NS}} = 0.005$ up to $\sigma_{\rm{BH}}/\sigma_{\rm{NS}} = 1.0$ (full NS kicks), as listed in column 2 of Table \ref{table:models}. Effectively, we use BH natal kicks \citep[a poorly constrained parameter; e.g.,][]{Mandel2016, Repetto2012} as a simple and convenient way to vary the number of BHs retained in our GC models. BH natal kicks are intended to serve as a proxy for more physical mechanisms that determine the numbers of BHs retained in GCs over long timescales.
 

\begin{deluxetable}{c|c||c|cc|c|c}
\tabletypesize{\scriptsize}
\tablewidth{0pt}
\tablecaption{Cluster properties for all model GCs\label{table:models}}
\tablehead{
	\colhead{Model} &
    \colhead{$\frac{\sigma_{\rm{BH}}}{\sigma_{\rm{NS}}}$} &
    \colhead{$M_{\rm{tot}}$}&
    \colhead{$r_c$}&
    \colhead{$r_h$}&
    \colhead{$N_{\rm{BH}}$}&
    \colhead{$N_{\rm{BH-MS}}$}\\
    \colhead{} &
    \colhead{} &
    \colhead{($10^5\,M_{\odot}$)}&
    \multicolumn{2}{c}{(\rm{pc})}&
    \colhead{}&
    \colhead{10 Gyr $<t<$ 12 Gyr}
}
\startdata
1 & 0.005 & 2.11 & 1.27 & 4.25 & 404 & 2 \\
2 & 0.01 & 2.1 & 1.97 & 4.46 & 424 & 2 \\
3 & 0.02 & 2.11 & 3.04 & 4.91 & 397 & 1 \\
4 & 0.03 & 2.09 & 3.77 & 4.91 & 389 & 0 \\
5 & 0.04 & 2.14 & 1.83 & 4.84 & 379 & 10 \\
6 & 0.05 & 2.12 & 3.13 & 4.12 & 383 & 0 \\
7 & 0.06 & 2.2 & 1.88 & 3.55 & 319 & 4 \\
8 & 0.07 & 2.16 & 1.64 & 4.43 & 285 & 1 \\
9 & 0.08 & 2.21 & 1.63 & 3.26 & 286 & 5 \\
10 & 0.09 & 2.22 & 2.23 & 2.8 & 249 & 2 \\
11 & 0.1 & 2.24 & 0.9 & 2.58 & 192 & 1 \\
12 & 0.2 & 2.31 & 0.11 & 1.64 & 7 & 9 \\
13 & 0.4 & 2.22 & 0.22 & 1.82 & 2 & 3 \\
14 & 0.6 & 2.33 & 0.21 & 1.78 & 4 & 0 \\
15 & 0.8 & 2.36 & 0.32 & 1.71 & 2 & 1 \\
16 & 1.0 & 2.3 & 0.21 & 1.56 & 3 & 2 \\
\enddata
\tablecomments{Column 2 shows the scaling of BH natal kicks, $\sigma_{\rm{BH}}/\sigma_{\rm{NS}}$, used for each model. Columns 3-6 show properties of each model at $t=12$ Gyr. Note that all models form $\sim 1500$ BHs initially. Column 6 shows the number of BHs that are retained at $t=12$ Gyr. Column 7 shows the number of different detached BH--MS binaries that appear in snapshots in the range 10 Gyr $<t<$ 12 Gyr, spanning the approximate uncertainty in the age of NGC 3201.}
\end{deluxetable}





\subsection{`Observing' model clusters}
\label{sec:obs_params}

It is not straightforward to compare GC models to observational data for a specific GC such as NGC\ 3201.

Each of our GC models contains $\sim100-1000$ snapshots in time spaced $\sim10-100$ Myr apart 
spanning the full evolution of the cluster, from formation to present day. From these snapshots, we create two-dimensional projections for each model at the corresponding snapshot times, assuming spherical symmetry. The half-light radius, $r_{\rm{hl}}$, is estimated by finding the projected radius which contains half of the cluster's total light. To estimate the core radius, $r_c$, we use the method described in \citet{Morscher2015} and \citet{Chatterjee2017}: we fit an analytic King model approximation \citep[Equation (18) of][]{King1962} to the cumulative stellar luminosity at a given projected radius including stars within a projected distance of $r_{\rm{hl}}$.

The methods used to calculate the surface brightness profile and velocity dispersion profile are described in detail in the Appendix.

\section{Results}
\label{sec:results}

\begin{figure}
\begin{center}
\includegraphics[scale=0.53]{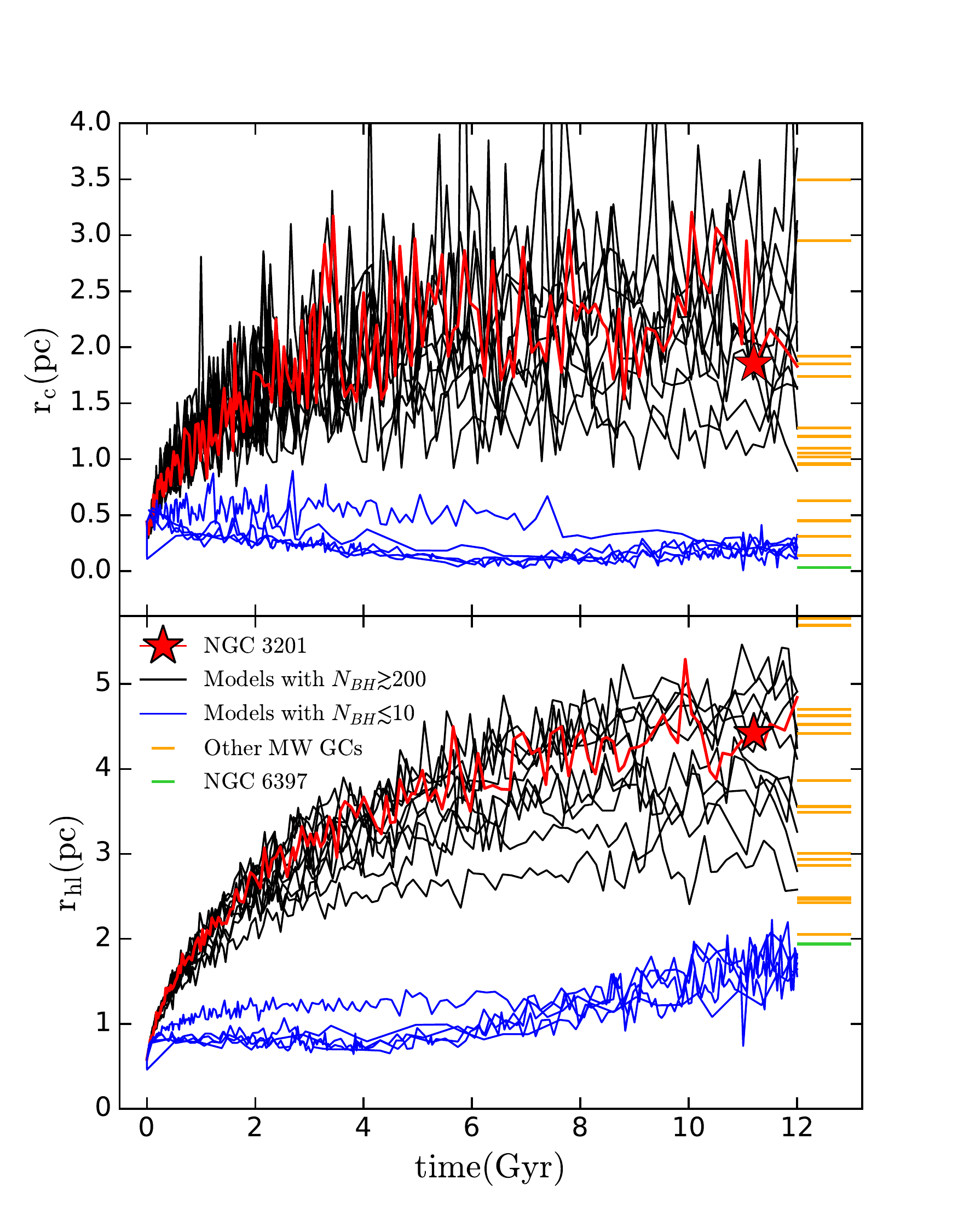}
\caption{\label{fig:rcrh} Core radius (top panel) and half-light radius (bottom panel) vs time for the models in Table \ref{table:models}. Black lines show models 1-11 (which retain $\gtrsim 200$ BHs at late times), blue show models 12-16 (which retain $\lesssim 10$ BHs at late times). The red curve marks model 5, our best-fit model. NGC 3201 is shown by the red star. The orange bars show the present-day properties of MW GCs similar to NGC 3201. The core-collapsed cluster NGC 6397 is shown as a green bar.}
\end{center}
\end{figure}

Figure \ref{fig:rcrh} shows the evolution of $r_c$ and $r_{\rm{hl}}$ for the 16 models in Table \ref{table:models}. Black lines show evolution of models 1-11, all of which retain $\gtrsim 200$ BHs at late times. Blue lines show models 12-16 which contain $\lesssim 10$ BHs at late times (see Table \ref{table:models}). Horizontal orange bars show $r_c$ and $r_{\rm{hl}}$ of MW GCs with absolute V-band magnitude and heliocentric distance similar to NGC 3201 (in the ranges of $-8.95$ to $-5.95$ magnitude and $6-12$ kpc, respectively; data taken from \citet{Harris1996}). As Figure \ref{fig:rcrh} shows, these MW GCs span a wide range in $r_c$ and $r_{\rm{hl}}$. As a limiting case, this set of MW GCs includes the core-collapsed cluster NGC 6397 (the green bar), which is shown to align with models containing very few BHs at present.

The red star marks the observed $r_c$ and $r_{\rm{hl}}$ for NGC 3201, which are consistent with models that retain large numbers of BHs.


The sustained heating and expansion shown by the red curves in Figure \ref{fig:rcrh} are primarily due to the frequent super-elastic interactions in the centrally-concentrated BH system: energy is deposited onto the stellar background when a recoiled BH sinks back via dynamical friction
\citep[see][]{Mackey2007,Mackey2008}.

\begin{figure}
\begin{center}
\includegraphics[width=1.0\columnwidth]{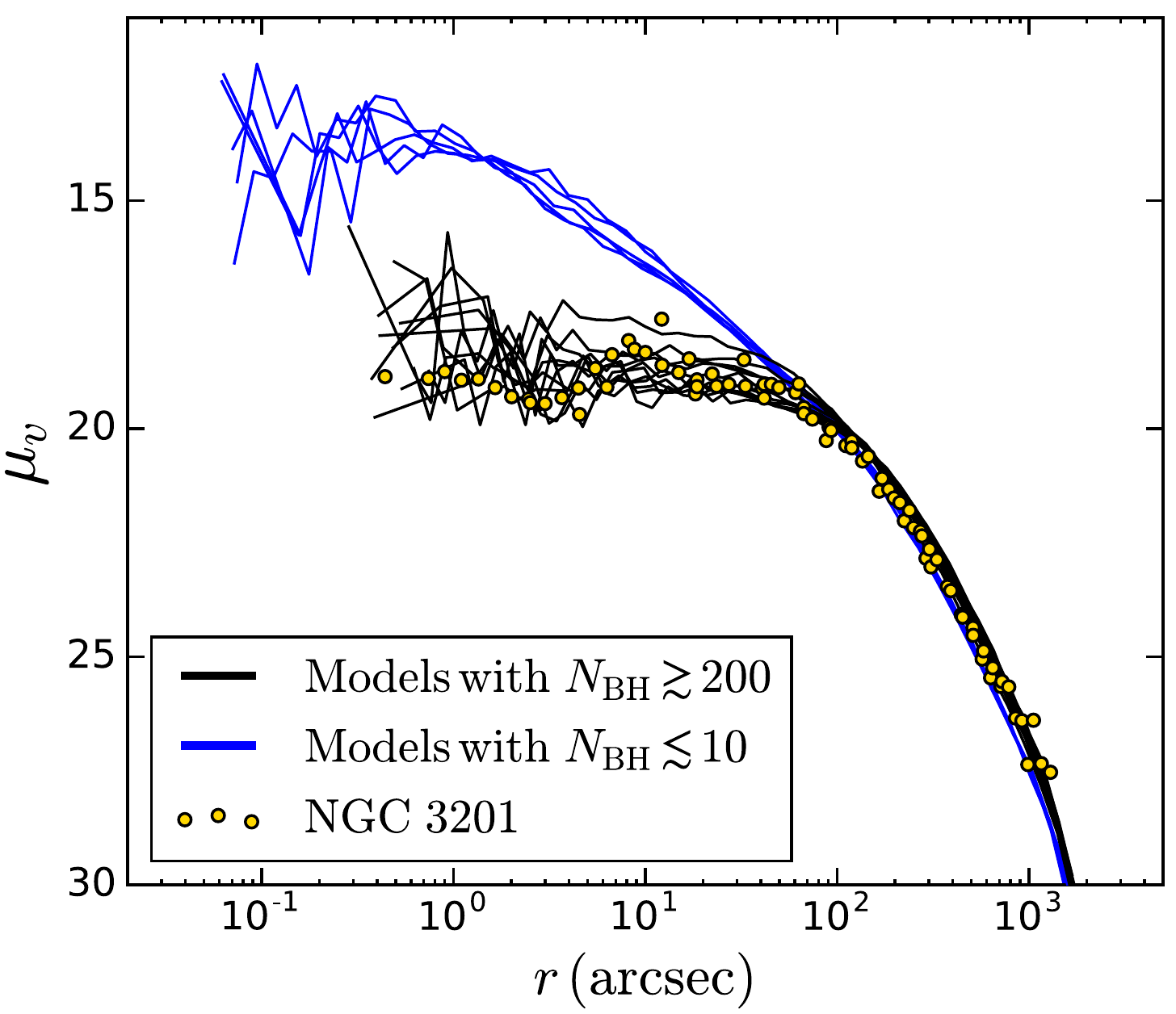}
\caption{\label{fig:SBP} Surface brightness profiles for all GC models at late times compared to NGC 3201 \citep[yellow circles; data from][]{Trager1995}. Black lines show models 1-11 (models retaining $\gtrsim 200$ BHs at late times) and blue lines show models 12-16 (models retaining $\lesssim 10$ BHs at late times; see Table \ref{table:models}.} 
\end{center}
\end{figure}

Figure \ref{fig:SBP} shows the surface brightness profile for each of the 16 models at $t=12$ Gyr (solid black and blue lines) compared to the observed surface brightness profile of NGC 3201 \citep[yellow circles;][]{Trager1995}. As in Figure \ref{fig:rcrh}, black lines show models 1-11 of Table \ref{table:models} and blue lines show models 12-16. 

As Figure \ref{fig:SBP} shows, models that retain many BHs have surface brightness profiles closest to matching NGC 3201. The surface brightness profiles of models retaining few BHs exhibit cusps at low $r$, representative of so-called ``core-collapsed'' MW GCs.

Henceforth, we use the term ``core-collapsed'' to simply denote those models (12-16) that remain much more centrally concentrated than their BH-heated counterparts (models 1-11), throughout their evolution, and have surface brightness profiles with prominent central cusps at late times as shown in Figure \ref{fig:SBP}.

The surface brightness profiles of all models converge with NGC 3201 for $r \gtrsim 100$ arcsec. This is simply because all models have similar total mass at late times (see column 3 of Table \ref{table:models}), which is a consequence of the choice of initial particle number ($N=8 \times 10^5$ for all models), and the same galactocentric distance (and hence tidal radius).

In order to identify a single best-fit model which accurately matches NGC 3201, we calculate the surface brightness profile, velocity dispersion profile, $r_c$, and $r_{\rm{hl}}$ for all snapshots at late times for each of the 16 models and compare to the observational data of NGC 3201. Note that several different estimates for the current age of NGC 3201 exist, ranging from $\sim10$ Gyr \citep{Forbes2010} up to $\sim12$ Gyr \citep{Usher2017}. To reflect this uncertainty in age, we simply consider all cluster snapshots in the range  10 Gyr $\leq t \leq$ 12 Gyr (the maximum evolution time considered in our grid of models).

We identify model 5 at $t=11.2$ Gyr as our best-fit model. For this model, $r_c=1.85$ pc and $r_{\rm{hl}}=4.42$ pc, which are in exact agreement with the $r_c$ and $r_{\rm{hl}}$ values for NGC 3201 given in \citet{Harris1996}. This particular model retains 389 BHs, including 4 detached BH--MS binaries. The assumptions that go into choosing our best-fit model are discussed in greater detail in the Appendix. 

\subsection{BH--MS binaries}
\label{sec:MS}

As shown in \citet{Chatterjee2017} and \citet{Kremer2018}, the formation of BH--non-BH binaries in a GC is a self-regulated process limited by a complex competition between the number of BHs and the number density of non-BH--non-BH binaries in the region of the GC where BHs mix with non-BHs. This competition ensures that the number of BH--non-BH binaries found in a GC at late times remains independent of the total number of BHs retained in the cluster.

All GC models considered here produce at most a few BH--MS binaries at late times (column 7 of Table \ref{table:models}), regardless of the total number of BHs retained at late times (column 6), in agreement with the results of \citet{Chatterjee2017} and \citet{Kremer2018}

In order to explore the formation of BH--MS binaries similar to the binary system recently detected in NGC 3201 \citep{Giesers2018}, we search for detached BH--MS binaries in models 1-11 (models that retain large numbers of BHs at late times and have structural parameters similar to those of NGC 3201; see Figure \ref{fig:SBP}).

\begin{figure}
\begin{center}
\includegraphics[width=1.0\columnwidth]{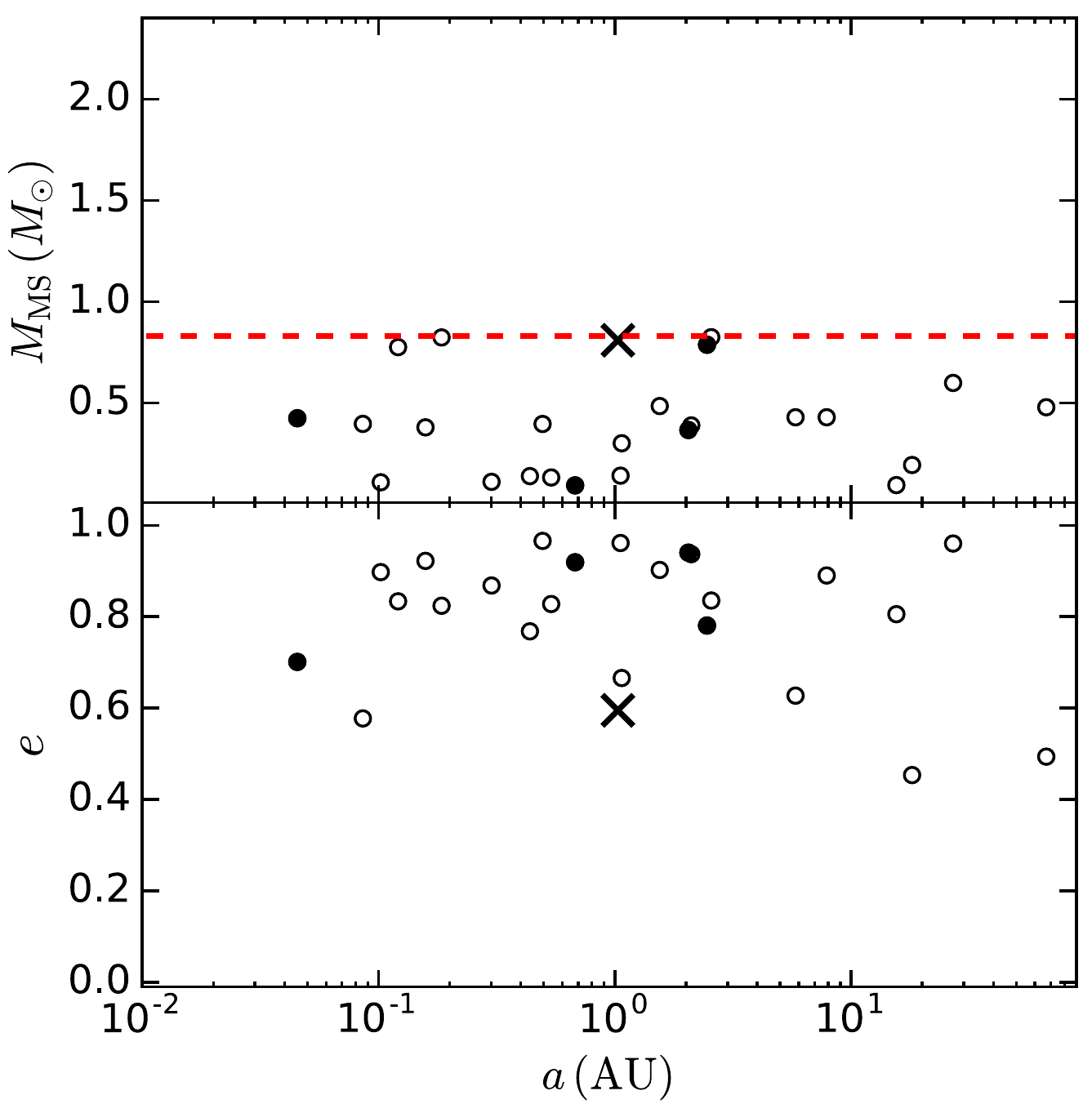}
\caption{\label{fig:scatter} Top panel (bottom panel) shows MS companion mass (eccentricity) vs semi-major axis for all detached BH--MS binaries found in NGC 3201-like models (models 1-11) at late times. Filled circles are the systems found in our best-fit model (model 5 at $t=11.2$ Gyr) and open circles show all systems in models 1-11. The black ``x'' marks the BH--MS binary observed in NGC 3201. The horizontal dashed line marks the turn-off mass.}
\end{center}
\end{figure}

In total, we find 24 different detached BH--MS binaries in models 1-11 in cluster snapshots between 10 Gyr $< t <$ 12 Gyr. The top panel of Figure \ref{fig:scatter} shows the MS companion mass, $M_{\rm{MS}}$, vs semi-major axis, $a$, for each of these binaries. Filled circles in Figure \ref{fig:scatter} mark the 4 BH--MS binaries found in model 5 at $t=11.2$ Gyr (our best-fit model described in the previous section). Open circles show BH--MS binaries found in all remaining snapshots in the range 10 Gyr $< t <$ 12 Gyr for models 1-11. The horizontal dashed line marks the turn-off mass ($M_{\rm{to}}=0.83 M_{\odot}$) for clusters of this particular age and metallicity. The black ``x'' marks the quoted values of the MS mass ($0.81\pm 0.05\,M_{\odot}$) and minimum $a$ ($1.03 \pm 0.03\,\rm{au}$) for the BH--MS binary detected in NGC 3201 \citep{Giesers2018}.

The bottom panel of Figure \ref{fig:scatter} shows the eccentricity, $e$, vs $a$ for all BH--MS found in models 1-11. Clearly, the majority of the detached BH--MS binaries identified at late times have moderate to high eccentricities, a consequence of the dynamical encounters experienced by these binaries.

Indeed, dynamical encounters play a critical role in the formation and evolution of these binaries. Of the 24 BH--MS binaries identified in these models, {\em all} are formed dynamically (none are primordial binaries). The median number of encounters experienced by the BHs in these binaries is 9. The median number of dynamical encounters experienced by the BH--MS binaries after formation, via an exchange encounter, is 2. Note that these BH--MS binaries are susceptible to destruction as the result of exchange encounters with other cluster objects. The median lifetime for the BH--MS binaries shown in Figure \ref{fig:scatter} is $\sim 400$ Myr.

The median BH mass for these 24 binaries is 9 $M_{\odot}$, consistent with the minimum mass estimated for the NGC 3201 BH-candidate  \citep[$M_{\rm{BH}} \sin(i) = 4.36 \pm 0.41 \, M_{\odot}$; ][]{Giesers2018}. A low BH-mass is anticipated based on our understanding of the evolution of BHs in GCs. The most massive BHs retained after formation will be ejected at early times through dynamical encounters. Only the least massive BHs will remain at late times \citep[e.g.,][]{Morscher2013}. Thus the detected BH--MS binary has properties that would be commonly produced dynamically in a GC similar to NGC\ 3201. 

Note that, in addition to the 24 detached BH--MS binaries, we also find 14 BH--MS binaries in mass-transferring configurations at late times, similar to those studied in \citet{Kremer2018}.

\subsection{Blue straggler--BH binaries}
\label{sec:BS}

\begin{figure}
\begin{center}
\includegraphics[width=1.0\columnwidth]{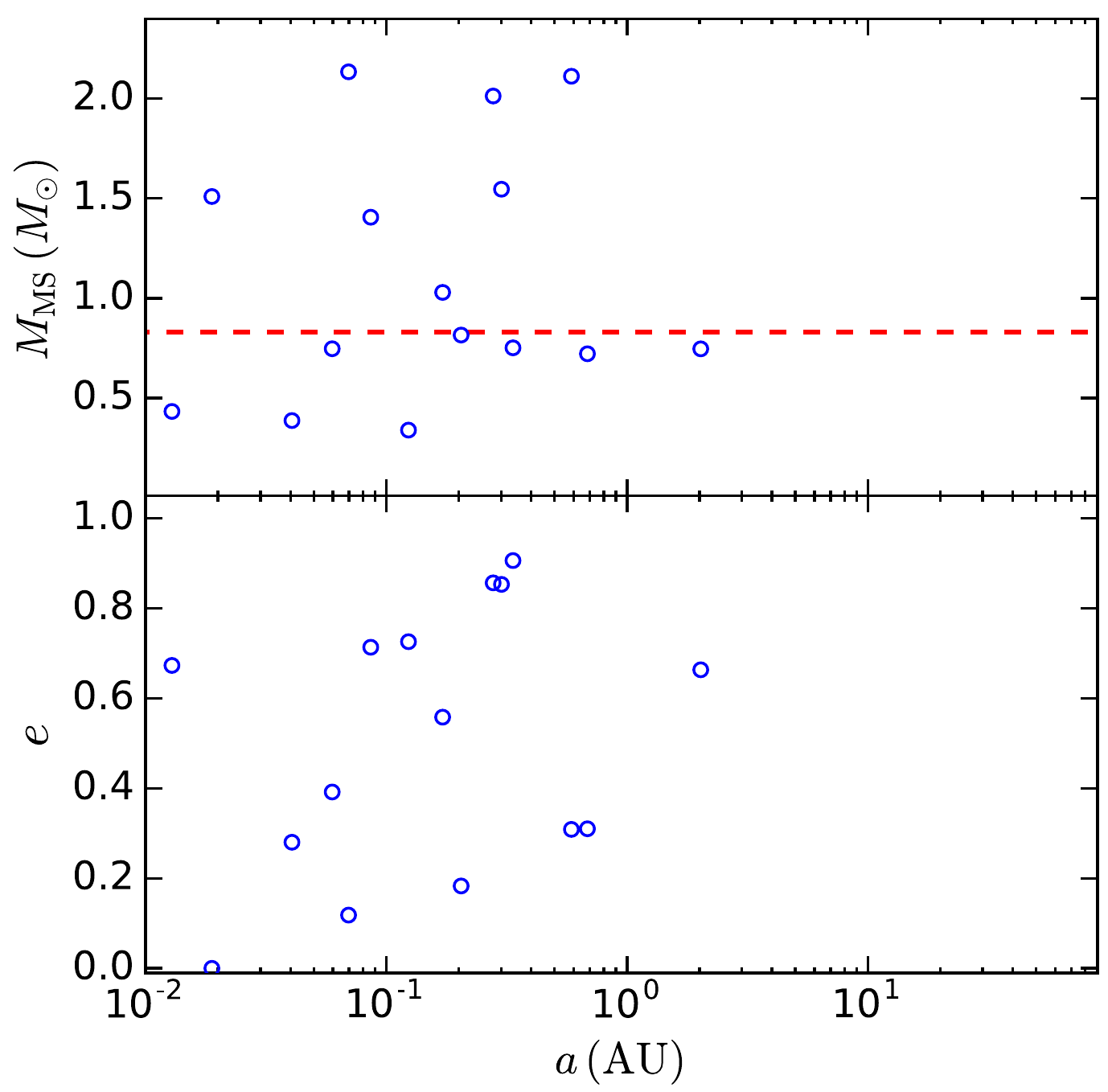}
\caption{\label{fig:BS} Analogous to Figure \ref{fig:scatter} but for all BH--MS systems found in models 12-16 (core-collapsed models which retain few BHs). Systems with $M_{\rm{MS}}$ above the horizontal dashed line (turn-off mass) are blue stragglers formed by collisions of MS stars which result from dynamical encounters.}
\end{center}
\end{figure}

Although models 12-16 are core-collapsed at late times and are therefore poor representations of NGC 3201, we show orbital parameters of all detached  BH--MS binaries found in these models in Figure \ref{fig:BS}. Interestingly, for these models, 
%
a large fraction of the MS star masses lie above the turn-off mass. These systems are blue stragglers (BSs) formed through collisions earlier in the evolution of their host cluster. Three of these BS companions have mass $M_{\rm{MS}} \gtrsim 2 \, M_{\odot}$, which is more than twice the turn-off mass. These are created via multiple collisions.

Among our models, the BH--BS binaries are unique to the core-collapsed clusters retaining few BHs. This is consistent with our understanding of GC evolution. As the number of retained BHs decreases, the central density increases, ultimately leading to core collapse. Additionally, in the absence of large numbers of retained BHs, the high-mass MS stars and MS binaries increasingly dominate the cluster's core, leading to an increased rate of dynamical interactions between multiple MS stars. Most of these interactions are resonant in nature and can often result in collisions, efficiently creating BSs \citep[e.g.,][]{Chatterjee2013b, Sills2013}. Being more massive than the average cluster object, the BSs readily interact and form binaries with the remaining BHs, like the systems seen in Figure \ref{fig:BS}. Such BH--BS binaries could be detectable through radial-velocity methods similar to those used to identify the BH candidate in NGC 3201. We do note that BSs have been observed in non-core-collapsed MW GCs, including NGC 3201 \citep{Simunovic2014}. We do find BSs in our BH-rich models, however none of these BSs are found with BH binary companions. We reserve a more detailed examination of BS formation in GCs for a future study.

Comparison of Figures \ref{fig:scatter} and \ref{fig:BS} show that the BH--MS binaries found in NGC 3201-like GCs and those found in core-collapsed clusters occupy significantly different regions of the parameter space. In addition to the presence of BH--BS binaries in core-collapsed models, the BH--MS binaries found in NGC 3201-like GCs tend to have higher eccentricities and are found over a wider range in $a$ (no BH--MS binaries with $a \gtrsim 10$ au are found in core-collapsed clusters). This is a direct consequence of higher central velocity dispersions in the more compact core-collapsed clusters which allows only very hard binaries to survive.

\section{Conclusion and Discussion}
\label{sec:discussion}

Using our Cluster Monte Carlo code, \texttt{CMC}, we have demonstrated a method to model the MW GC NGC 3201. We showed that by varying the magnitude of BH natal kicks,  the number of retained BHs at late times changes substantially, which in turn, determines the observational features of our GC models at late times. In particular, we showed that in order to produce a GC model that matches the observational properties (including surface brightness profile, velocity dispersion profile, core radius, and half-light radius) of NGC 3201 at present, the model must retain $\gtrsim 200$ BHs.

Additionally, we demonstrated that GC models retaining large numbers of BHs at late times can harbor detached BH--MS binaries similar in properties to the system recently detected through radial-velocity measurements in NGC 3201.

We also showed that although GCs which retain few BHs are poor representations of NGC 3201, they may serve as good representations of core-collapsed MW GCs. 
We demonstrated that such core-collapsed GCs may contain bright blue straggler--BH binaries, that could be detectable using radial-velocity methods similar to those used to identify the BH--MS system in NGC 3201.

Previous analyses have modeled other GCs with known stellar-mass BH candidates. For example, \citet{Heggie2014} (Monte Carlo) and \citet{Sippel2012} (N-body) explored models of M22, and demonstrated similar results to those of this letter; namely, that retained BHs provide an energy source capable of producing models that match GCs with large observed core radii at late times and that a fraction of the BHs remaining at late times are likely to be found in binaries with luminous companions.

Additionally, previous analyses have shown that, in the absence of BHs, binary hardening may provide sufficient energy to roughly balance the energy diffusion from two-body relaxation, after contraction of the cluster's core. This allows the cluster to attain a quasi-steady state even when the binary fraction is only a few percent \citep{Chatterjee2013a, Heggie2003, Vesperini1994}.

\citet{Chatterjee2013a} showed how clusters that have entered this ``binary-burning'' phase exhibit cuspy surface brightness profiles typical of the observed core-collapsed MW GCs and similar to GC models presented here that retain few BHs. The binary fraction for NGC 3201 is estimated to be low \citep[$\sim 5-15$\%;][]{Cote1994}, consistent with our choice of initial binary fraction. This suggests that large numbers of BHs must be retained to produce a cluster similar to NGC 3201 since models with too few retained BHs would lead to surface brightness profiles vastly different from that of NGC 3201. See the Appendix for further discussion of the possible effects of binaries.

In this study we used BH natal kicks as a proxy for adjusting the number of BHs retained in GCs at late times. In reality, other factors likely determine whether or not a particular GC retains large numbers of BHs at present. In particular, initial cluster parameters such as virial radius, cluster mass, binary fraction, concentration, and galactocentric distance determine how dynamically evolved a GC is at present. This in turn may determine how efficiently BHs are dynamically ejected from GCs via recoil from dynamical encounters. We will explore effects of these other initial parameters in future work. 


While we consider only BH--MS binaries in this paper, populations of other classes of compact objects may also correlate with GC evolution, and thus, the retained  BH population. In particular, the number of neutron stars (NSs) in GCs that can be observed as X-ray binaries (NS--XRBs) or as millisecond pulsars may correlate with the number of BHs retained at late times. We note that no pulsars or NS--XRBs have been observed in NGC 3201,  unlike, for example, the core-collapsed cluster NGC 6397. We also find that NS-XRBs and millisecond pulsars rarely form in our models retaining many BHs (NGC 3201-like models), but instead we find these objects form readily in our core-collapsed models retaining few BHs.





\acknowledgments
We thank the referee for helpful comments and suggestions.

This work was supported by NASA ATP Grant NNX14AP92G 
and NSF Grant AST-1716762. K.K. acknowledges support by the National Science Foundation Graduate Research Fellowship Program under Grant No. DGE-1324585.
S.C. acknowledges support from
CIERA, the National Aeronautics and Space Administration
through a Chandra Award Number TM5-16004X/NAS8-
03060 issued by the Chandra X-ray Observatory Center
(operated by the Smithsonian Astrophysical Observatory for and on behalf of the National Aeronautics
Space Administration under contract NAS8-03060), 
and Hubble Space Telescope Archival research 
grant HST-AR-14555.001-A (from the Space Telescope 
Science Institute, which is operated by the Association of Universities for Research in Astronomy, Incorporated, under NASA contract NAS5-26555).

\listofchanges

\appendix
 
\section{Calculating surface brightness and velocity dispersion profiles} 
\label{sec:best-fit}

\begin{figure}
\begin{center}
\includegraphics[width=0.5\linewidth]{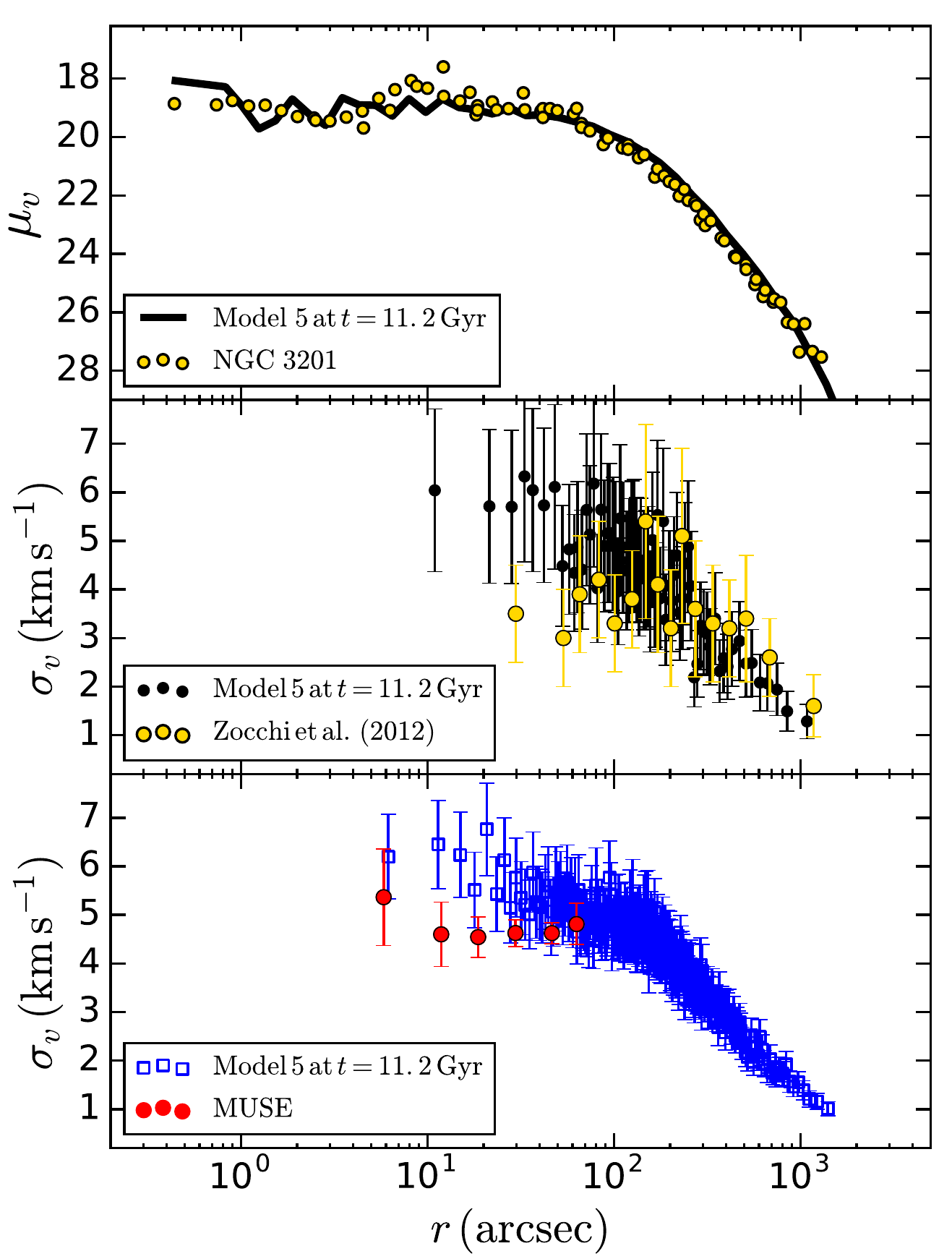}
\caption{\label{fig:bestfit} Surface brightness profile (top) and velocity dispersion profiles (middle and bottom) for model 5 at $t=11.2$ Gyr, which we identify as our best-fit model for NGC 3201. In the middle panel, we compare to the observed velocity dispersion profile of \citet{Zocchi2012}, which considers only giants, and in the bottom panel we compare to observations from the MUSE survey \citep{Kamann2017}, which considers main-sequence stars and giants.
}
\end{center}
\end{figure}

Figure \ref{fig:bestfit} shows the surface brightness profile (top panel) and velocity dispersion profile (bottom two panels) of our best-fit model: model 5 in Table \ref{table:models} at $t=11.2$ Gyr.  Here we describe our method for calculating surface brightness profiles and velocity dispersion profiles for our models at different points in time.

To calculate the surface brightness profile of a model GC, we produce two-dimensional projections of each time snapshot as described in Section \ref{sec:obs_params}, then divide these two-dimensional projections into 50 equally spaced radial bins. To calculate the surface brightness within each bin, we remove all stars with luminosity $L_{\star} > 15 \, L_{\odot}$.  We compare to the observed surface brightness profile from \citet{Trager1995}. Note that the central density (inside the core radius) of our best-fit model in physical units is $4.4 \times 10^3\,\rm{pc}^{-3}$.

To calculate the velocity dispersion profile, we use the method of \citet{Pryor1993} to find the velocity dispersion within a set of radial bins. We compare to two observational profiles (shown here with $2\sigma$ error bars): \citet{Zocchi2012}, which calculates the velocity dispersion from observations of giants using radial velocity data from \citet{Cote1995}, and the \texttt{MUSE} survey \citep{Kamann2017}, which uses observations of main sequence stars and giants to calculate the dispersion. To compare to \citet{Zocchi2012} (middle panel of Figure \ref{fig:bestfit}), we include only giants and group into radial bins of 25 stars each. For comparison to MUSE, we include all main-sequence and giant stars with $M \geq 0.4 M_{\odot}$ and choose radial bins such that each bin contains at least 100 stars and covered an annulus of $\log(r/\rm{arcsec}) \geq 0.2$, as in \citet{Kamann2017}.

Because the positions and velocities of stars in our models can change significantly from snapshot to snapshot, and as a consequence of the low number of stars per bin, the calculated value of $\sigma_v$ will fluctuate between snapshots, particularly at small $r$. Nevertheless, as Figure \ref{fig:bestfit} shows, the velocity dispersion profile of our best-fit model falls within the $2\sigma$ error bars of the observations.

Uncertainty in several observational features of NGC 3201 at present, in particular the cluster's age and distance from Earth, give us freedom to explore ranges in these parameters in order to find a model that accurately matches the observational data of NGC 3201, including the surface brightness and velocity dispersion profiles, as well as $r_c$ and $r_{\rm{hl}}$ (calculated as described in Section \ref{sec:obs_params}).

The uncertainty in the age of NGC 3201 \citep[e.g.,][]{Forbes2010, Usher2017} allows us to treat all cluster snapshots with ages in range 10 Gyr $<t<$ 12 Gyr as equally valid representations of NGC 3201 at present. Generally, the total cluster mass and total number of retained BHs change by only $\sim 5-10$\% within this time range, however, as shown in Figure \ref{fig:rcrh}, $r_c$ and $r_{\rm{hl}}$ can fluctuate significantly between successive snapshots. As a result, by exploring different time snapshots we can identify models with $r_c$ and $r_{\rm{hl}}$ most similar to NGC 3201.

Because the total cluster mass changes by only small amounts in the time range 10--12 Gyr, the outer parts of the surface brightness and velocity dispersion profiles ($r \gtrsim 10^2$ arcsec), which are most sensitive to the total cluster mass, do not change significantly between snapshots. However, in the innermost parts ($r \lesssim 10^2$ arcsec), both the surface brightness and velocity dispersion profiles fluctuate between snapshots because of the increased error due to low $N$ in these regions. 

Secondly, the uncertainty in the distance to NGC 3201 \citep[$d_{\odot}=5.0 \pm 0.4$ kpc;][]{Covino1997}, allows the surface brightness profiles for our models to be shifted by small amounts horizontally in either direction which alter the fit to the observational data. For our best-fit model, we adopt a distance of 5.7 kpc, within $2\sigma$ of the published distance to the cluster.

\begin{figure}
\begin{center}
\includegraphics[width=0.5\linewidth]{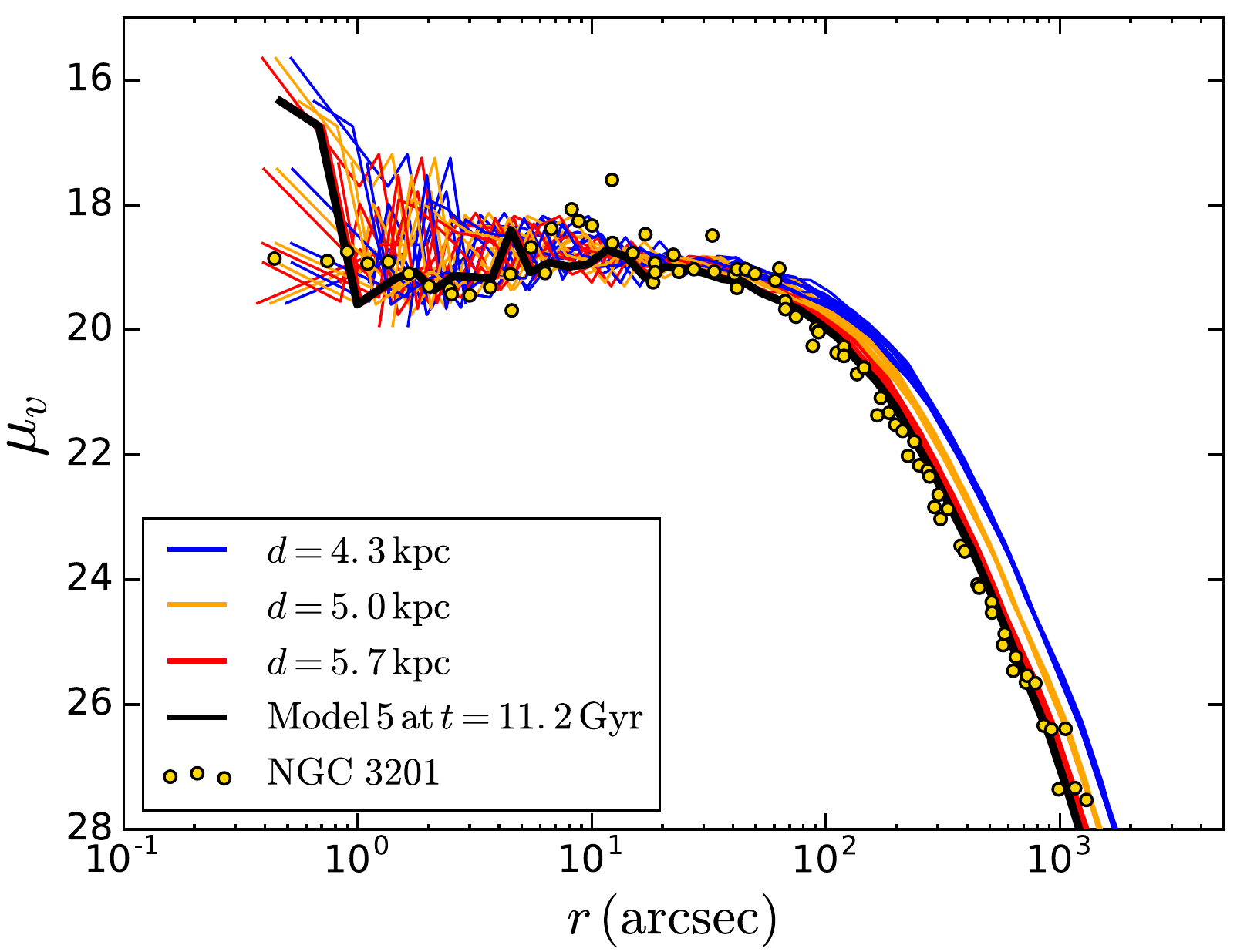}
\caption{\label{fig:SBP_error} Surface brightness profiles for all late-time snapshots for model 5 assuming three different distances: 4.3 kpc (blue), 5 kpc (orange), and 5.7 kpc (red). Our best-fit model (model 5 at $t=11.2$ Gyr) is shown in black.}
\end{center}
\end{figure}

Figure \ref{fig:SBP_error} shows how the calculated surface brightness profile for model 5 (our best-fit model) changes based on our choice of cluster age and distance from Earth. Here, the blue plots show surface brightness profiles for all snapshots with 10 Gyr $<t<$ 12 Gyr assuming a distance of 4.3 kpc. Orange plots show profiles for all snapshots assuming a distance of 5 kpc. Red shows the same for a distance of 5.7 kpc, which is distance adopted for our best-fit model, which is shown in the figure as the black curve.

As the figure shows, at small $r$, changes in snapshot time are the dominant effect in the variations in the surface brightness profile, while at large $r$, the choice of distance is the dominant effect.

\section{Binary fraction}

As Figure \ref{fig:SBP} demonstrates, with all other initial 
cluster parameters fixed, the scaling of BH natal kicks, and, as a consequence, the number of BHs retained, has a substantial effect on the observational characteristics of a GC model at late times. However, it is unclear from the grid of models used in this study whether variation of other initial parameters may also have similar effects. In particular, previous studies have shown that binaries may provide a sufficient energy source to prevent deep core-collapse, and in absence of sufficient energy production from BH dynamics, energy can be extracted via binary-hardening leading to a quasi-stedy core the size of which depends on the binary fraction \citep[e.g.,][]{Chatterjee2013a,Vesperini1994}.

For the grid of models in this analysis, we choose an initial binary fraction of 5\%, consistent with the observed binary fraction of NGC 3201 \citep[$\sim 5-15\%$;][]{Cote1994}. However, we also ran a single model identical to model 16 in Table \ref{table:models} (with $\sigma_{\rm{BH}}=\sigma_{\rm{NS}}$), but with an initial binary fraction of 50\%. At $t=12$ Gyr, this model retains few BHs ($N_{\rm{BH}}=3$) and is core-collapsed. This suggests that even with a binary fraction significantly higher than that suggested by observations, BHs are likely necessary to reproduce the observational features of NGC 3201.

\section{Virial radius}

In this analysis, we assume an initial virial radius of 1 pc, but we also ran a second grid of models identical to those listed in Table \ref{table:models}, but with $r_v=2$ pc. Initially larger clusters are less dynamically evolved at a fixed present-day age. Thus, in general, these models retain $\sim 10\%$ more BHs than their $r_v=1$ pc counterparts, but the trend shown in Figure \ref{fig:SBP} still holds: models which retain few BHs at late times are core-collapsed, while models with many BHs have observational features similar to NGC 3201. Taking advantage of the uncertainty in distance and age for NGC 3201, as described in Section \ref{sec:best-fit}, we can similarly identify a best-fit model from the $r_v=2$pc grid, and draw similar conclusions to those drawn in this analysis. In particular, an NGC 3201-like model must retain several hundred BHs, and such a model can naturally produce BH--MS binaries similar to the BH candidate recently observed in NGC 3201 composed of a low-mass BH and a MS star close in mass to the turn-off.

\end{document}